\documentclass{elsart}
\usepackage{amsmath}
\usepackage{amssymb}
\usepackage{graphicx}
\newcommand{\V}{\mathcal{V}}
\newcommand{\U}{\mathcal{U}}
\newcommand{\ins}{\text{ins}}
\newcommand{\del}{\text{del}}

\DeclareMathOperator{\Div}{\mathrm{div}}
\usepackage{natbib}

\begin{document}

\begin{frontmatter}
  \title{Simple Mathematical Model Of Pathologic Microsatellite
    Expansions: When Self-Reparation Does Not Work}
\author{Boris Veytsman}

\address{George Mason University, School of Computational Sciences,
  MS~5C3, Fairfax, VA 22030, USA, E-mail: \texttt{borisv@lk.net}}

\author{Leila  Akhmadeyeva}

\address{Bashkir State Medical University, 3 Lenina Str.,
    Ufa, 450077, Russia.  E-mail: \texttt{medic@agidel.ru}}

\begin{abstract}
  We propose a simple model of pathologic microsatellite expansion,
  and describe an inherent self-repairing mechanism working against
  expansion.  We prove that if the probabilities of elementary
  expansions and contractions are equal, microsatellite expansions are
  always self-repairing.  If these probabilities are different,
  self-reparation does not work.  Mosaicism, anticipation and reverse
  mutation cases are discussed in the framework of the model.  We
  explain these phenomena and provide some theoretical evidence for
  their properties, for example the rarity of reverse mutations.
\end{abstract}


\end{frontmatter}

\section{Introduction}
\label{sec:intro}

Pathologic microsatellite expansion is a phenomenon causing several
severe diseases like Fragile~X, Huntington disease, Myotonic Dystrophy
and others~\citep{Harper01:MDBook, Pearson03:_slipping_while_sleeping,
  Pineiro03, Mockton03-sca7tg, Pearson05Review}.  There are places in
a DNA molecule where nucleotide sequences are repeated several times.
The number of such repeats (satellites) is usually stable during
normal replication.  However, sometimes a mutation occurs, and the
mutated DNA has more (expansion) or less (contraction) repeats than
its ancestor.  Normally the mutation rates are about
$10^{-3}\dots10^{-4}$ per generation per
locus~\citep{EllegrenReview00}.  However in the case of diseases
mentioned above, the expansions occur much faster, at a rate of
hundreds or thousands per locus per generation.  We will call this
phenomenon \emph{pathologic expansion} to distinguish it from the much
slower ``normal'' expansion.  We are interested in the case where the
number of nucleotides in the repeating sequence is small, e.g.  when
the repeated sequences are triplets.  In this case the phenomenon is
usually called microsatellite expansion.  Sometimes, as in the case of
Myotonic Dystrophy Type 1 (DM1, OMIM~\#160900), the number of repeated
triplets might actually reach thousands.  For some diseases this
expansion occurs in a coding part of DNA, for some in a non-coding
one, but it is always a multi-system disease with multiple symptoms.

There are several notable features of pathologic microsatellite
expansion, common to most diseases associated with it:
\begin{description}
\item[Mosaicism:] For most diseases the number of repeats is
  \emph{not} the same in all cells.  Rather, it has a wide
  distribution of possible values. A notable exception is Huntington
  disease (HD, OMIM~\#143100), where the mosaicism is not as prominent
  as for other expansion-related diseases~\citep{Harper02:HD}.  The
  reason for becomes more clear after we discuss the model.  We return
  to this disease in Section~\ref{sec:concl}.
\item[Anticipation:] For some diseases a relatively small increase in
  the number of repeats does not lead to symptoms.  However, the
  stability of the repeated sequence is lower than the stability of
  the non-affected DNA, and the children of affected parents 
  might show symptoms, sometimes severe~\citep{Harper01:MDBook}. 
\item[Reverse Mutation:] Sometimes the children of symptomatic
  patients have the normal number of repeats.  This is a rare, but still
  observable phenomenon~\citep{Brunner93:ReverseMD, Monckton95-dmsppcrhmg}.
\end{description}
A theory of microsatellite expansion must naturally explain these
phenomena.

One of the most common explanations of microsatellite expansion is the
formation of hairpins either during
replication~\citep{Cleary02:_cis_factors,
  Mirkin02:_positioned_to_expand, Pearson03:_slipping_while_sleeping,
  yang03:_replic_inhib_modul_instab_of} or during DNA repair after
transcription~\citep{Monckton04-pms2}.  In both these cases the
hairpin formation can cause either expansion or contraction of DNA.  A
recent review comparing these explanations can be found in,
e.g.~\citep{Pearson05Review}.  In this paper we will not try to
distinguish between these mechanisms; the proposed model describes
both.  Therefore we will understand by \emph{cell events} either cell
divisions in the first model or cell repair events in the second one.

This model is attractive because it can explain a number of features
of microsatellite expansion.  In this model hairpins form during some,
but not all, cell events.  Therefore it is a random, rather than a
deterministic process.  Thus different cells have different number of
expansions and contractions in their histories.  This explains
mosaicism, i.e. broad distribution of the number of repeats in
different cells within the same tissue.

Gametes in this model might have different numbers of repeats.  If the
number of repeats turns out to be small, the child of an affected
parent will be not affected.  This explains reverse mutation.  The
calculations below show that this phenomenon is indeed very rare.

To explain anticipation, we can assume that the probability of
expansion grows with the number of repeats in the DNA molecule.  If
the number of repeats in an asymptomatic patient increases, the
probability to have a symptomatic child also increases.

One of the ways to verify these speculations is to try to make
possible conclusions from the model, and to check whether these
conclusions agree with the observed picture of microsatellite
expansion.  If they do, our confidence in the model grows, if they do
not, then the model is wrong.  This paper takes the qualitative model
described above for granted and tries to formalize it in the form of
differential equations for the observed distribution of repeats.  We
solve these equations and show a qualitative agreement with the
observations.

It is interesting to compare the pathologic microsatellite expansion
due to fast mutations with the ``normal'' microsatellite expansion due
to slow mutations.  The latter got much attention (see
e.g.~\citep{DiRienze94, FitzSimmons95, Goldstein95, Angers97,
  Primmer98, Schloetterer98, Brinkmann98, Makova00, Kayser00,
  EllegrenNature00, Huang02, Calabrese03} and the
review~\citep{EllegrenReview00}) as a way to infer data on evolution
process.  This approach has a promise of higher time resolution than
other methods because ``slow'' microsatellite mutations are still
several orders of magnitude faster than most other
mutations~\citep{EllegrenReview00}.  Such full comparison of
``normal'' and pathologic expansions is beyond this paper, but one
might express a hope that in the future it will help to understand
both better.  For example, while during ``normal'' mutations the
expansions of the repeat sequence are thought to be more probable than
the contractions, there seems to exist some mechanism that limits an
uncontrolled expansion of microsatellites~\citep{Garza95, Amos96,
  Harr00, Xu00}.  Apparently such mechanism is absent or too week for
pathologic microsatellite expansions.  A comparative study might
therefore help to elucidate details and effects of this mechanism.  On
the other hand, one must be very cautious in the application of data
and conclusions from the study of the ``normal'' expansions to the
pathologic ones and vice versa.

There are many theoretical works describing ``normal'' microsatellite
expansion using both analytical methods and computer simulations (see
e.g.~\citep{Tachida92, Shriver93, Nielsen97, Bell97, Kruglyak98,
  Kruglyak00, Calabrese01, Sibly01, Whittaker03, Shinde03, Sibly03,
  Calabrese03, Lai03a, Lai03, Lai03b}).  However, they deal with
slowly changing sequences with relatively small number of repeats.
Our situation is rather opposite: we are interested in  long sequences
and fast change.  Therefore our formalism and results are quite
different from theirs.

\section{Model}
\label{sec:model}

First, let us discuss how one hairpin is formed.  Consider a hairpin
with $h/2$ repeats.  If $l_k$ is the number of repeats in a Kuhn
segment of the polymer~\citep{ScalingConcepts, FundamentalsPlmsc},
then we gain $h/(2l_k)$ degrees of freedom.  The corresponding free
energy loss is $kTh/l_k$, where $k$ is Boltzmann constant, $T$ is
temperature.  On the other hand the energy gain is $\delta E h$, where
$\delta E$ is the energy gain per repeat.  Since both these
contributions are proportional to $h$, the total free energy change is
also proportional to $h$:
\begin{equation}
  \label{eq:Delta_F}
  \Delta F = Ch,\qquad C=\text{const}
\end{equation}
The value of the constant $C$ in this equation depends on the relation
between $\delta E$ and $kT/l_k$.  If $C<0$, the formation of hairpins
causes a decrease of free energy, and the longer are the hairpins, the
better.  This would lead to a fast de-stabilization of the number of
repeats.  Since this does not happen, we can conclude that $C>0$.
This means that the formation of hairpins is \emph{not} encouraged by
thermodynamics, and the formation of longer hairpins is suppressed
with the probability proportional to $\exp(-Ch)$.  Since the
probability exponentially decreases with $h$, only the shortest
possible hairpins are formed.  The minimal size of a hairpin depends
on the flexibility of the molecule.  It stands to reason to assume it
of the order of one-two Kuhn segments.  Therefore the microsatellite
de-stabilization cannot start until the DNA has at least several $l_k$
repeats.

These thermodynamic considerations explain anticipation: it is
necessary to have at least several Kuhn segments in the microsatellite
repeats interval to start the mechanism of de-stabilization.  Of
course \emph{cis}-elements might subtly influence hairpin formation at
the early stages of de-stabilization.  Therefore they play an
important role in the transition from anticipation to
disease~\citep{Monckton99-Cis, Cleary02:_cis_factors}.  It is
interesting that a certain threshold number of repeats is necessary
for ``normal'' expansions too, at least in some cases~\citep{Sibly01,
  Sibly03, Shinde03}. 

Let us now discuss a strand of DNA having $x$ repeats after $i$ cell
events.  The next event can have one of three possible outcomes:
\begin{enumerate}
\item No expansion or contraction occurred.
\item There was an expansion of length $n$.  Let
  $Q_{\text{ins}}(x,n)$ be the probability of this event.
\item There was a contraction of length $n$. Let
  $Q_{\text{del}}(x,n)$ be the probability of this event.
\end{enumerate}
Let $P_i(x)$ be the probability that the strand has exactly $x$
repeats.  Then it is easy to write the master equation, describing the
transition from the step number $i$  to the step number $i+1$:
\begin{multline}
  \label{eq:master}
  P_{i+1}(x) = P_i(x) + \\
  \sum_{n=1}^\infty \Bigl( P_i(x-n)Q_{\ins}(x-n,n) + 
    P_i(x+n)Q_{\del}(x+n,n) -\\
    P_i(x)Q_{\ins}(x,n) - P_i(x)Q_{\del}(x,n)\Bigr)
\end{multline}
Equation~(\ref{eq:master}) might be simplified if we make the
following assumption, based on the thermodynamic considerations in the
beginning of this Section.  Namely, we assume that the constant $C$ in
equation~(\ref{eq:Delta_F}) is large enough, so expansions and
contractions are in fact rare.  If $n_{\min}$ is the minimal hairpin
length allowed by chain flexibility, then the only events to be
considered in the sum~(\ref{eq:master}) are expansions and
contractions of length $n_{\min}$.  Now we must estimate the
probabilities of one expansion or contraction as functions of repeats
number $x$.  If we consider for guidance ``slow'' mutations in the non
pathologic regime (see Introduction), we see that there is a
considerable controversy in the literature about the dependence of
mutation rate on $x$.  Some authors report exponential
growth~\citep{Brinkmann98, Whittaker03, Lai03}, while other report
much weaker linear relationship~\citep{Kruglyak98, Kruglyak00,
  Sibly01, Shinde03} or more complex dependence~\citep{Calabrese03,
  Sibly03}.  Moreover, \emph{cis}-factors, obviously, should also
influence the mutation rates.  We can only agree with
\citet{Primmer98a}: ``These observations demonstrate that the mutation
process of microsatellites may be more complex than previously
thought''.  Fortunately the situation for relatively large $x$ can be
simplified.  Indeed, if the number of repeats is sufficiently large,
we can divide the stretch of microsatellites into parts, each enough
to assume that a mutation or a repair error in one part does not
affect the other ones.  Only two end parts depend on the
\emph{cis}-factors.  If each part mutates independently, the overall
mutation rate should be proportional to the number of parts.  This
simple consideration suggests that the mutation rates at least for
large $x$ should be linear in $x$.  Moreover, they must go to zero as
the number of repeats goes to $n_{\min}$.  If we set the origin of
$x$ axis to $n_{\min}$, we get simply:
\begin{equation}
  \label{eq:Q}
  Q_{\ins}(x,n) = q_{\ins}(n) x,\qquad
  Q_{\del}(x,n) = q_{\del}(n) x
\end{equation}
With these assumptions equation~\eqref{eq:master} can be rewritten as:
\begin{multline}
  \label{eq:master-simp2}
  P_{i+1}(x) - P_i(x) =  \\
  q_{\ins}(n_{\min}) \Bigl(
  P_i(x-n_{\min})(x-n_{\min}) -
  P_i(x)x
  \Bigr) +\\
  q_{\del}(n_{\min}) \Bigl(
  P_i(x+n_{\min})(x+n_{\min}) -
  P_i(x)x
  \Bigr)
\end{multline}

The next step is the transition from the discrete
representation~(\ref{eq:master-simp2}) to a continuous one.  We will
``smooth'' the variables $i$ and $x$.  In order to do this we will
measure ``time'' $t$ in the number of events and consider $P$ to be a
function of a continuous variables $t$ and $x$, so $P(t,x)\d x$ is the
probability to have the number of repeats between $x$ and $x+dx$ at
the time $t$.  Then we can rewrite equation~(\ref{eq:master-simp2}) in
the continuous form as:
\begin{equation}
  \label{eq:diffusion}
  \begin{gathered}
    \frac{\partial P(t,x)}{\partial t} = 
    -c \frac{\partial \bigl(xP(t,x)\bigr)}{\partial x}
    + D\frac{\partial^2
      \bigl(xP(t,x)\bigr)}{\partial x^2}\\
    c=\bigl(q_{\ins}(n_{\min})-q_{\del}(n_{\min})\bigr) n_{\min}\\
    D=\frac{q_{\ins}(n_{\min})+q_{\del}(n_{\min})}{2} n_{\min}^2
  \end{gathered}
\end{equation}
Note that the quantity
\begin{equation}
  \label{eq:J}
  J = cxP -D\frac{\partial (x P)}{\partial x}
\end{equation}
has the meaning of \emph{flow} of probability through the point $x$ at
the time $t$. By the way, this means that
equation~\eqref{eq:diffusion} has the simple meaning of continuity
equation $\partial P/\partial t + \Div J=0$. 

If the number of repeats in the zygote is $x_0$, then
equation~(\ref{eq:diffusion}) has the following initial condition:
\begin{equation}
  \label{eq:P_init}
  P(0,x) = \delta(x-x_0)
\end{equation}
where $\delta$ is Dirac's delta-function~\citep[e.g.][]{GreenFuncs}.

We will see that $P(t,+0)$ remains finite, so
\begin{equation}
  \label{eq:P_border}
 \lim_{x\to0} xP(t,x)=0
\end{equation}
As shown below, the flow~\eqref{eq:J} remains non-zero at $x\to+0$.
Therefore the integral
\begin{equation}
  \label{eq:f_m}
  f_m(t) = \int_{+0}^{\infty} P(t,x)\d x 
\end{equation}
is \emph{not} conserved. This integral represents the fraction of
``mutant'' cells, i.e. cells with the number of repeats large enough
to form hairpins and therefore to be described by
equation~\eqref{eq:diffusion}.  The discontinuity of the function
$P(t,x)$ at $x\to0$ makes this integral less than 1.  Its complement
to 1 is the fraction of cells, which can no longer form hairpins and
are ``stuck at zero'':
\begin{equation}
  \label{eq:f_r}
  f_r(t) =1-f_m(t)
\end{equation}
We will call such cells ``repaired'' cells.  The increase of $f_r$
over time represents a self-reparation effect.

We introduce the parameter
\begin{equation}
  \label{eq:gamma}
  \gamma = \frac{c x_0}{D}
\end{equation}
This parameter reflects the difference between the probabilities of
expansion and contraction.  The case of $\gamma=0$ corresponds to the
situation when expansions and contractions occur with equal
probabilities.  If expansions are more probable, then $\gamma>0$.
Note that the value of $\gamma$ depends on the progenitor number of
repeats $x_0$.  The greater is $x_0$, the larger is $\gamma$.  We will
see that this parameter critically affects the microsatellite
instability.

We will measure time in the units of $x_0/D$, i.e. we will introduce a
dimensionless variable
\begin{equation}
  \label{eq:tau}
  \tau = tD/x_0
\end{equation}
As shown in Appendix, at the reasonable values for the parameters one
dimensionless unit of time corresponds to about 25 cell events.

\section{Results And Discussion}
\label{sec:results}

The solution for equation~\eqref{eq:diffusion} is obtained in
Appendix~\ref{sec:solution}.  Here we discuss the properties of the
solution and predictions of the model.

First we consider the fraction of repaired cells (see
Appendix~\ref{sec:solution}): 
\begin{equation}
  \label{eq:f_r-sol}
  f_r(\tau) = 
  \exp\left(-\frac{\gamma}{1-\e^{-\gamma\tau}}\right)
\end{equation}
The number of ``repaired'' cells increases with the time $\tau$.  The
speed of this increase and the limit fraction at $\tau\to\infty$
depend on the parameter $\gamma$.

In the special case $\gamma=0$, i.e. when expansions and contractions
happen with the same probability, equation~\eqref{eq:f_r-sol} becomes
$f_r(\tau) = \exp\left(-1/\tau\right)$.  In this case $f_r$ goes to 1
as $\tau\to\infty$.  This means that all cells eventually become
repaired.

In the case $\gamma>0$ the limit $f_r$ at $\tau\to\infty$ is
$\exp(-\gamma)$.  For large enough $\gamma$ the fraction of repaired
cells is small, but nevertheless not zero.  This case corresponds to
the observed clinical picture.

Now we can explain the phenomenon of reverse mutation.  If a parent is
affected, the gamete might carry DNA either from the repaired
population or from unrepaired, mutant population.  In the first case a
reverse mutation occurs.  Therefore the probability of reverse
mutation is $\exp(-\gamma)$.  It seems that reverse mutations are very
rare events.  In the case of Myotonic
Dystrophy~\citep{Brunner93:ReverseMD} the probability of reverse
mutation is very small.  We will rather arbitrarily estimate it as
$1:1000$; a more frequent occurrence would be observed more often, and
a more rare one would not be observed at all.  This gives the
following estimate for $\gamma$:
\begin{equation}
  \label{eq:estimate_gamma}
  \gamma\approx7
\end{equation}

\begin{figure}
  \includegraphics{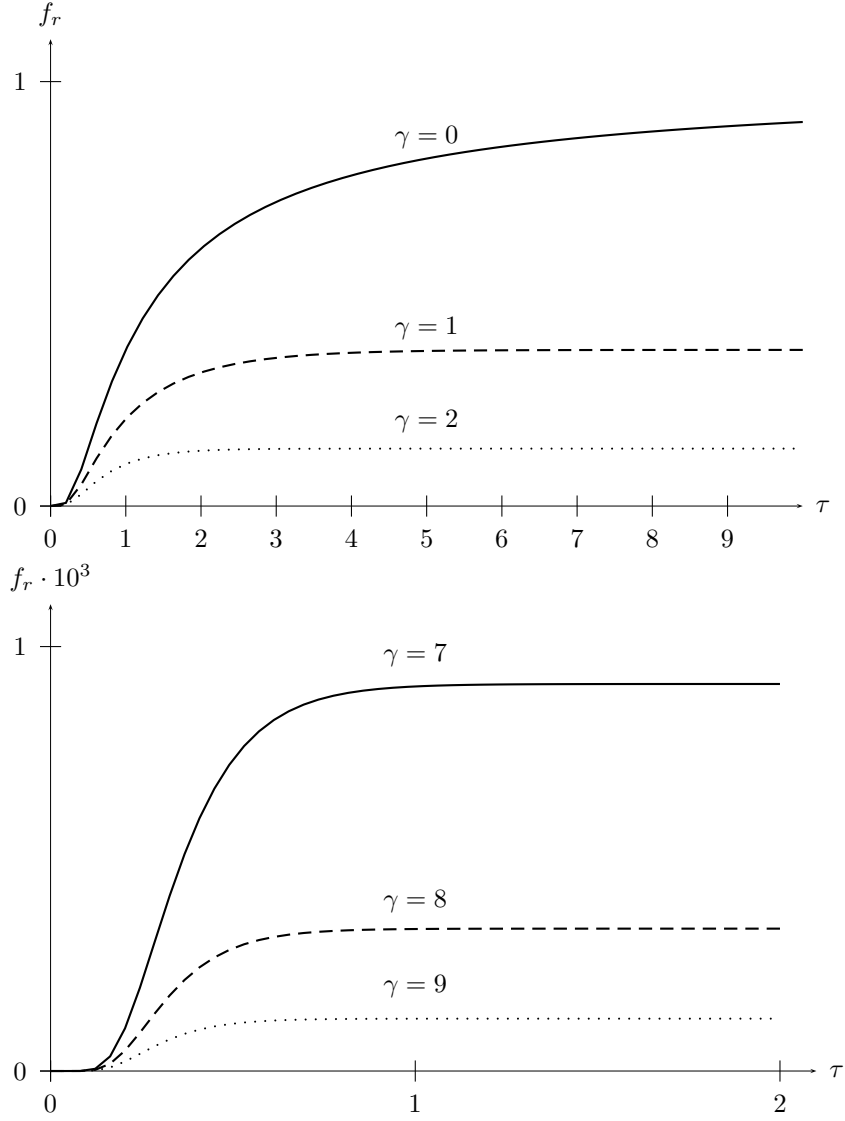}
  \caption{Fraction of Repaired Cells As Function of Time}
  \label{fig:f_r(tau)}
\end{figure}

Plots of $f_r(\tau)$ for several values of $\gamma$ are shown on
Figure~\ref{fig:f_r(tau)}.  It can be seen from these plots that the
number of repaired cells quickly reaches the limit value.  This
justifies the assumption that the fraction of repaired cells in the
gametes is equal to the limiting value.

Let us now return to the solution of equation~\eqref{eq:diffusion}. It
can be expressed through the mean number of repeats $m$ and standard
deviation $\sigma$ (see Appendix~\ref{sec:solution}): 
\begin{equation}
  \label{eq:P-sol}
  P(x,t) = \frac{2m^{3/2}}{\sigma^2 x^{1/2}} 
  \exp\left[
      -\frac{2 m^2}{\sigma^2}\left(
        1+\frac{x}{m}
      \right)
    \right]
    I_1\left(
      \frac{4m^{3/2}x^{1/2}}{\sigma^2}
    \right)
\end{equation}
where $I_1$ is the modified Bessel
function~\citep[\S~9]{AbramowitzStegun}.  The mean number of repeats
and standard deviation depend on time as
\begin{equation}
  \label{eq:m-sigma}
  m=x_0\exp(\gamma\tau),\qquad
  \sigma =
   \left(
    2\,\frac{1-\exp(-\gamma\tau)}{\gamma}
    \right)^{1/2}
    x_0  \exp(\gamma\tau) 
\end{equation}
Also interesting are skewness and kurtosis of the distribution.  They
are
\begin{equation}
  \label{eq:S-K}
  S = 3 \left(
    \frac{1-\exp(-\gamma\tau)}{2\gamma}
    \right)^{1/2},\qquad
  K = 6\,\frac{1-\exp(-\gamma\tau)}{\gamma}
\end{equation}

At early stage of instability growth ($\gamma\tau\ll1$) these
equations describe a sharp distribution centered around $m$.  The
ratio of the distribution width $2\sigma$ to the mean size $m$ is small
(about $2(2\tau)^{1/2}$).

However, at later stages ($\gamma\tau\gg1$) the picture is completely
different.  At these stages the curve is very wide.  The ratio of the
width to the the mean size is at these stages $2\sigma/m =
2(2/\gamma)^{1/2}\approx 1$.  This large width explains the observed
mosaicism.  

The transition from the first regime to the second one depends on
$\gamma$, and thus on the progenitor number of repeats $x_0$.  The
larger is $x_0$, the earlier is the transition to the second regime,
i.e. the regime of developed instability. 

The distribution has positive skewness and kurtosis.  They are about
zero at early stages, and tend to $3/(2\gamma)^{1/2}\approx 0.8$ and
$6/\gamma\approx 0.9$ correspondingly.

\begin{figure}
  \includegraphics{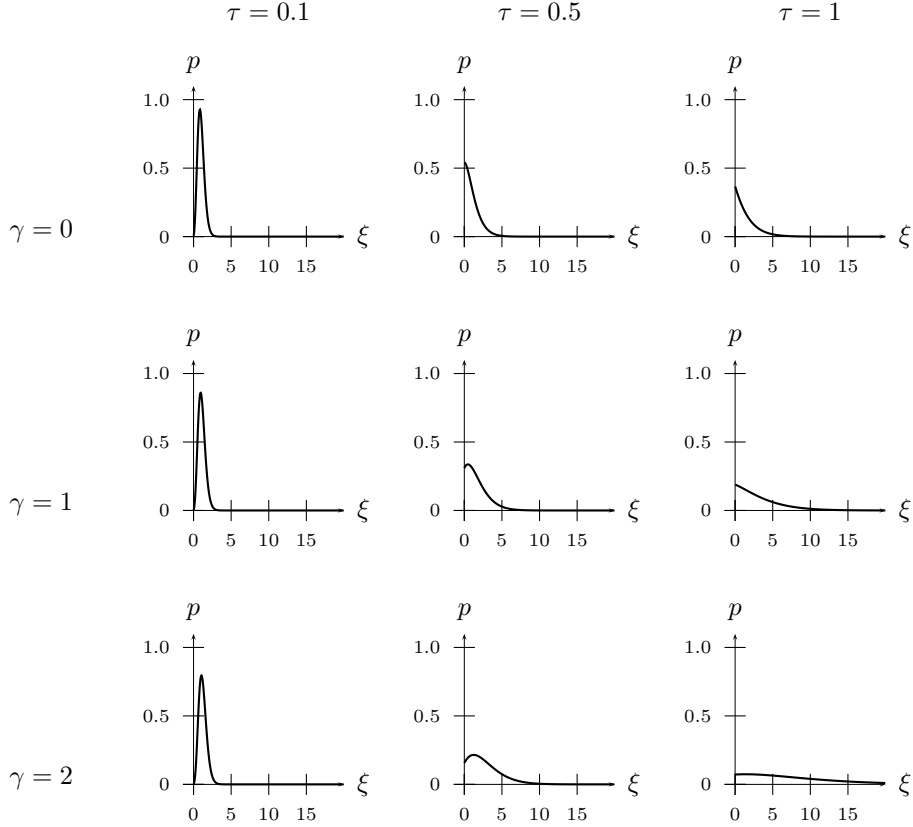}
    \caption{Repeat Number Distribution For Unrepaired Cells, small $\gamma$}
    \label{fig:size-distrib}
\end{figure}

\begin{figure}
  \includegraphics{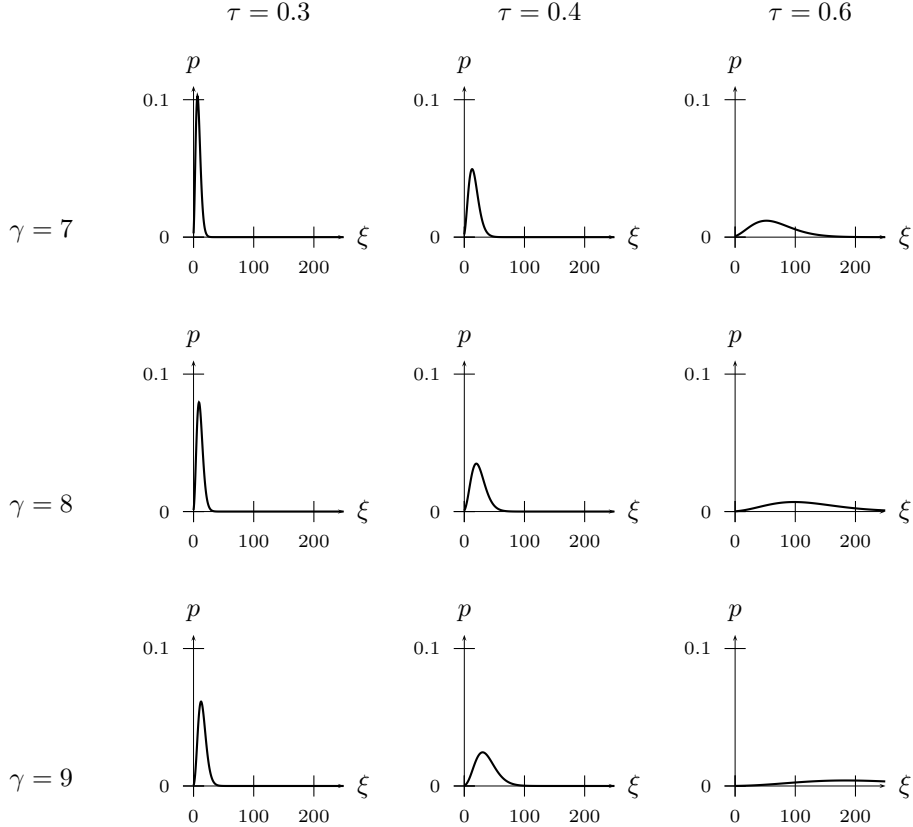}
    \caption{Repeat Number Distribution For Unrepaired Cells, large $\gamma$}
    \label{fig:size-distrib-large}
\end{figure}

Typical plots distribution of repeat lengths  are shown on
Figures~\ref{fig:size-distrib} and~\ref{fig:size-distrib-large}.

\section{Conclusions}
\label{sec:concl}

We have shown that a very simple model of pathologic microsatellite
expansion can qualitatively explain the observed phenomena of
anticipation, mosaicism and spontaneous recovery.  This model
considers expansion or contraction of repeats as a random process with
the probability of expansion and contraction related to the
probability of hairpin formation.  A mathematical model based on this
picture is able to predict the shape of the distribution of the number
of repeats after many divisions.  

This model predicts a natural ``reparation process'' leading to
reverse mutation.  In the case when the probabilities of expansion and
contraction are equal, this process eventually heals the mutation.
Therefore mutation survives only if the probability of expansion
exceeds the probability of contraction.  The fraction of repaired cells
in the long run depends on this difference in probabilities.

We implicitly assumed that only ``young'' cells, the ones belonging to
the latest generation, are used in the measurements of the number of
repeats.  If this assumption is not satisfied, the observed length
distribution should be obtained by summation of the results over
generations of cells.  However, if the ``older'' cells die due to
apoptosis, this effect is small.  For example, if the apoptosis for
blood cells occurs after 25--40 mitoses, as it is usually thought,
then the effect is indeed negligible for blood samples.

Another interesting question is the possibility of selection:  the
rate of cell survival and multiplication might depend on the level of
the mutation of microsatellite expansion.  This will change the rates
of expansion and contraction for the cell population as a whole.

A notable exception from the general picture of trinucleotide
expansion diseases is Huntington disease.  Mosaicism for this disease
is not as prominent as for other dynamic
mutations~\citep{Harper02:HD}. However, a closer look shows that this
example actually does not contradict our model.  The number of repeats
for HD is rather small (several dozens).  It seems that the mutation
in this case is caused by a small number of relatively large
expansions, rather than a large number of small expansions, as assumed
in this paper.

It would be interesting to extend the analysis of this paper to
quantitative comparison with the experimental data.  This will be done
in subsequent works.

A further comparison of the fast pathologic mutations and slow
``normal'' ones seems also to be promising.

\section*{Acknowledgements}
\label{sec:ack}

The authors are grateful to R.~Colby, C.E.~Pearson, R.~Korneluk,
K.~Makova, D.~Monckton, Y.~Giguere and F.~Rousseau for helpful
discussions.  This work was supported by International Research and
Exchange Board (IREX, 2003 ECA Alumni Small Grant), International
Association for the Promotion of Cooperation with Scientists from the
New Independent States of the Former Soviet Union (INTAS, grant~YSF
00-34), Russia President Grant Program for Young Scientists
(grant~MD-346.2003.04), Open Society Institute (Soros Foundation) and
American Austrian Foundation (AAF).

%\clearpage

\appendix

\section{Solution Of Master Equation}
\label{sec:solution}

In this Appendix we provide the solution of master
equation~\eqref{eq:diffusion}.  This equation is easier to solve in
the following dimensionless variables:
\begin{equation}
  \label{eq:dimensionless}
  \xi=x/x_0,\qquad \tau = tD/x_0, \qquad p(\tau,\xi)=x_0 P(x,t)
\end{equation}
Let us roughly estimate these parameters.  Taking values approximating
the known data about DM1, we get
\begin{equation}
  \label{eq:estimates_parameters}
  x_0\approx10^2,\qquad
  q_{\ins}\approx q_{\del}\approx10^{-2},\qquad
  n_{\min}\approx20
\end{equation}
so
\begin{equation}
  \label{eq:estimate_tau}
  \xi\approx10^{-2}x,\qquad
  \tau\approx0.04t
\end{equation}
In other words, one dimensionless unit of $\tau$ corresponds
approximately to 25 cell events, while one dimensionless unit of $\xi$
corresponds approximately to 100 repeats.

Let us introduce the function
\begin{equation}
  \label{eq:v}
  v(\tau,\xi)=\xi p(\tau,\xi)
\end{equation}
Then equations~\eqref{eq:diffusion} can be rewritten as
\begin{equation}
  \label{eq:diff-v}
      \frac{\partial v}{\partial \tau} 
      + \gamma\xi\frac{\partial v}{\partial\xi} - 
    \xi\frac{\partial^2 v}{\partial \xi^2} = 0
\end{equation}
We use Laplace transform with respect to $\xi$:
\begin{equation}
  \label{eq:Laplace}
  \V(\tau,s) = \int_0^{\infty} \e^{-s\xi} v(\tau,x)\d\xi
\end{equation}
For the Laplace transform of the derivatives we have
(see~\citep[\S~29.2.5]{AbramowitzStegun}) 
\begin{equation}
  \label{eq:derivatives}
    \frac{\partial v}{\partial \xi} \fallingdotseq
    s\V - v(t,+0),\qquad
    \frac{\partial^2 v}{\partial \xi^2} \fallingdotseq
    s^2\V - sv(t,+0) - V(\tau) 
\end{equation}
where $\fallingdotseq$ means Laplace transform, and
\begin{equation}
  \label{eq:V}
  V(\tau)=\left.\frac{\partial v(\tau,\xi)}{\partial
  \xi}\right\rvert_{\xi\to+0} 
\end{equation}

Multiplication by $-\xi$ corresponds to differentiation by $s$
(see~\citep[\S~29.2.9]{AbramowitzStegun}).  Due to the border
condition~\eqref{eq:P_border}, the values of $v(t,+0)$ and $V(\tau)$
in equation~\eqref{eq:derivatives} do not depend on $\xi$.  Therefore
we can rewrite equation~\eqref{eq:diff-v} in Laplace space as
\begin{equation}
  \label{eq:diffusion-laplace}
  \frac{\partial \V}{\partial \tau} + 
  \frac{\partial }{\partial s}\left(s^2\V -\gamma s \V\right) = 0  
\end{equation}
with the initial condition from equation~\eqref{eq:P_init}
\begin{equation}
  \label{eq:init-laplace}
  \V(0,s) = \exp(-s)
\end{equation}

Let
\begin{equation}
  \label{eq:diffeq-u}
  \U(\tau,s)=(s^2-\gamma s)\V(\tau,s)
\end{equation}
Then we can rewrite equations~\eqref{eq:diffusion-laplace}
and~\eqref{eq:init-laplace} as
\begin{gather}
  \label{eq:eq-u}
    \frac{\partial \U}{\partial \tau} + 
  (s^2-\gamma s)\frac{\partial \U}{\partial s} = 0  \\
  \label{eq:border-u}
  \U(0,s) = (s^2-\gamma s) \exp(-s)
\end{gather}

The solution of differential
equation~\eqref{eq:diffeq-u} is
\begin{equation}
  \label{eq:U-sol1}
  \U(\tau,s) = f\left(\frac{\e^{-\gamma\tau}(s-\gamma)}{s}\right)
\end{equation}
Taking into account the initial condition~\eqref{eq:border-u} and
returning to the function $\V$, we get
\begin{equation}
  \label{eq:V-sol}
  \V(\tau,s) = \frac{\e^{-\gamma\tau}\gamma^2}{\left(s(1-\e^{-\gamma\tau})+\e^{-\gamma\tau}\gamma\right)^2} 
  \exp\left(-\frac{s\gamma}{s(1-\e^{-\gamma\tau})+\e^{-\gamma\tau}\gamma}\right)
\end{equation}
The reverse Laplace transform of this expression is
(see~\citep[\S~29.3.81]{AbramowitzStegun}) 
\begin{equation}
  \label{eq:v-sol}
  v=\frac{\gamma\left(\e^{-\gamma\tau}\xi\right)^{1/2}}{1-\e^{-\gamma\tau}}
  \exp\left(-\frac{\gamma(1+\e^{-\gamma\tau}\xi)}{1-\e^{-\gamma\tau}}\right)
  I_1\left[\frac{2\gamma}{1-\e^{-\gamma\tau}}
    \left(\e^{-\gamma\tau}\xi\right)^{1/2}
  \right]
\end{equation}
which provides the solution of the master equation.

Using the
asymptotic~\citep[\S~9.6.7]{AbramowitzStegun}
\begin{equation}
  \label{eq:bessel_small_z}
  I_1(z)\approx z/2,\qquad z\ll1
\end{equation}
we see that at $\xi\to0$
\begin{equation}
  \label{eq:p(0)}
  p(\tau,+0)=\frac{\gamma^2\e^{-\gamma\tau}}{(1-\e^{-\gamma\tau})^2}\,
  \exp\left(-\frac{\gamma}{1-\e^{-\gamma\tau}}\right)
\end{equation}
The flow at $\xi\to+0$ is non-zero.  The total fraction of repaired
cells can be calculated by calculating the integral of flow.
Rewriting equation~\eqref{eq:J} in dimensionless coordinates, we see
that 
\begin{equation}
  \label{eq:f_r1}
  f_r(\tau) = \int_0^{\tau} p(u,0)\,du= 
  \exp\left(-\frac{\gamma}{1-\e^{-\gamma\tau}}\right)
\end{equation}
which gives equation~\eqref{eq:f_r-sol}.

The momenta of function $p(\xi)$ are defined as
\begin{equation}
  \label{eq:mu_d_n}
  \mu'_n(\tau) = \int_0^{\infty} \xi^n p(\tau,\xi)\d\xi,\qquad
  n=0, 1, 2,\dots
\end{equation}
Laplace transform gives
\begin{equation}
  \label{eq:mu_d_n-lim}
  \mu'_n(\tau) = (-1)^{n-1}\lim_{s\to+0} \frac{\partial^{n-1}
    \V(\tau,s)}{\partial s^{n-1}}, \qquad n=1,2,\dots
\end{equation}
Differentiating the function $\V$, calculating central momenta
(see~\citep[\S~26]{AbramowitzStegun}) and returning to dimensional
coordinates, we obtain equations~\eqref{eq:m-sigma} and~\eqref{eq:S-K}
for mean, deviation, skewness and kurtosis.  After transformation of
equation~\eqref{eq:v-sol} to dimensional coordinates and substitution
of equations~\eqref{eq:m-sigma}, we obtain equation~\eqref{eq:P-sol}.

%\clearpage

%\bibliography{neurology,polymers,physmath,hbonds,satdiffusion,genetics}
%\bibliographystyle{plainnat}

\end{document}